# Polyimide-Based Flexible Coupled-Coils Design and Load-Shift Keying Analysis

Yuan Yao, Wing-Hung Ki and Chi-Ying Tsui
Department of Electronic and Computer Engineering
The Hong Kong University of Science and Technology
Hong Kong SAR, P. R. China
yyaoah@ust.hk, eeki@ust.hk, eetsui@ust.hk

*Abstract*—Wireless power transfer using inductive coupling is commonly used for medical implantable device. The design of the secondary coil on the implantable device is important as it will affect the power transfer efficiency, the size of the implant and also the data transmission between the implant and the in-vitro controller. In this paper, we present a design of the secondary coil on polyimide-based flexible substrate to achieve high power transfer efficiency. Load shift keying modulation is used for the data communication between the primary and secondary coils. Thorough analysis is done for the ideal and practical scenario and it shows that a mismatched secondary LC tank will affect the communication range and communication correctnes. A solution to achieve robust data transmission is proposed and then verified by SPICE simulations.

## I. Introduction

Implantable medical devices (IMD) usually have a high constraint on the size, and thus wireless power transfer is widely used instead of the bulky batteries to provide power for the implants. A wirelessly powered trans-sclera electrical stimulation (TsES) system has been developed to treat retinal degenerative diseases [1]. However, the power transfer efficiency of the inductive link was only 8% at a distance of 10 mm, which may not adequate for *in vivo* operations. The secondary coil was fabricated on a printed circuit board (PCB), which was too rigid to be used inside the body. Moreover, implant biocompatibility could not be guaranteed. In this paper, to solve the above problem, a pair of polyimide-based flexible coupled-coils is proposed to solve the above problems.

To save the use of an additional antenna, power and data are transmitted via the same inductive coupling link. Frequency-shift keying (FSK) modulation is commonly used in IMDs applications to achieve high data rate [2], but using FSK will increase circuit complexity, and high data rate is not necessary for many applications. To minimize the system volume, load-shift keying (LSK) modulation is also commonly used [3]-[5]. However, it was found that the uplink data sent back from the implant to the transmitter is not as robust as expected [1], limiting the communication range. In this paper, the issue of using LSK scheme is studied and analyzed. A solution is then suggested and verified by SPICE simulations.

## II. Polyimide-Based Flexible Coupled-Coils Design

IMDs are usually low-power applications and inductive power transfer using a pair of series-parallel (S-P) resonant coupled-coils is usually used. In [1] and [6], for an trans-sclera electrical stimulation system, the secondary coil is fabricated on a PCB with thickness of 0.3 mm, and after packaging with encapsulation, the implant thickness will be up to 1 mm.

**Table I Proposed Coupled-Coils Parameters**

|  | **Primary Coil** | **Secondary Coil** |
|---|---|---|
| Outer Length $l$ | N/A | 14 mm |
| Self-inductance $L$ | 895 nH | 564 nH |
| Series resistance $R_s$ | 1.114 Ω | 2.333 Ω |
| Quality factor @ 40.68 $MHz$ | 205.33 | 61.9 |

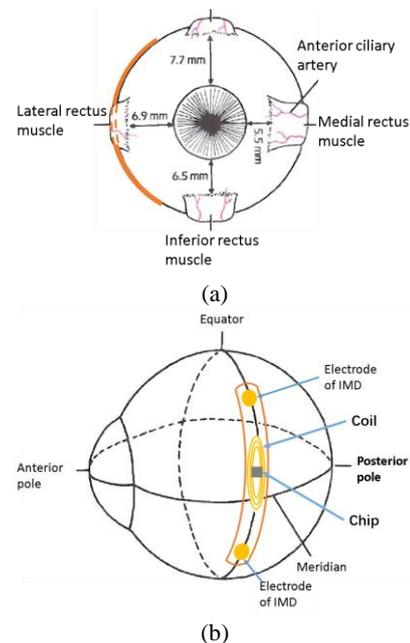

(a)

(b)

Fig. 1. Proposed secondary coil installation [9]

Surgeons suggest to put it beneath the conjunctiva and rectus muscle, and the implant width should be less than 6 mm so as not to affect the activity of the eyeball.

Polyimide has been proven to be a biocompatible material and an excellent choice for neuroprosthetic application [7], [8]. In this research the polyimide-based flexible secondary coil has been developed to overcome the size limitation and improve the power transfer efficiency (PTE). The proposed secondary coil installation is shown in Fig. 1.

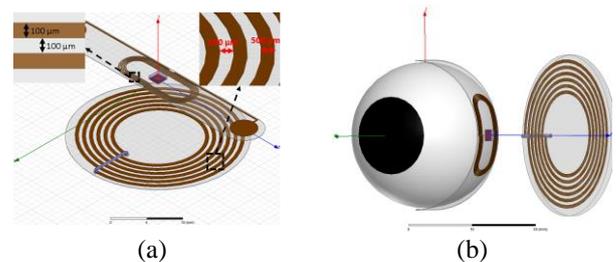

(a)      (b)

Fig. 2. Polyimide-based flexible coupled coils simulation setup (a) flat case. (b) bending case

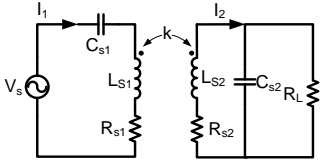

Fig. 3. Circuit model of inductive link

The two electrodes and coil are fabricated on the polyimide substrate, and the controller IC was mounted on the substrate as shown in Fig. 1. Proposed secondary coil installation [9]. As the substrate is biocompatible, the biocompatible packaging is easier to achieve, and the implant thickness is reduced to 0.2 mm. With the flexible substrate, the implant can be attached to the surface of the eyeball so that the implant can be made larger without affecting the eyeball movement. The width is still kept at 6 mm, while the length is increased to 40 mm for better stimulation performance. The shape of the coil is oval to make better use of the area.

To increase the link efficiency, the primary coil could also be made on a flexible PCB, such that it can be attached onto the skin to minimize the distance between the coils. The secondary coil is bended and attached to the surface of the eyeball. Bending will change both the self-inductances and the coupling coefficient, and the field solver HFSS is used to simulate the coupled-coils under the bending situation. The simulation setup is shown in Fig. 2. Parameters of the coupled-coils are shown in Table I. The primary has series resonance and the secondary has parallel resonance. The inductive link circuit model is shown in Fig. 3.

From [10], the inductor and capacitor values under the resonant condition is related by the following equation:

$$\omega_o = \frac{1}{\sqrt{L_{s2}C_{s2}}}\sqrt{1 - \frac{L_{s2}}{C_{s2}R_L^2}} \quad (1)$$

and the PTE is given by

$$\eta = \frac{P_{out}}{P_{in}} = \frac{k^2 Q_1 Q_2^2}{(1 + \frac{Q_2}{\alpha} + k^2 Q_1 Q_2)(\alpha + Q_2)} \quad (2)$$

$$Q_1 = \frac{\omega_o L_{s1}}{R_{s1}}, Q_2 = \frac{\omega_o L_{s2}}{R_{s2}}, \alpha \equiv \omega_o C_{s2} R_L \quad (3)$$

When both the primary and secondary were flat and 10 mm apart, the simulated PTE of the inductive link reached 39.33%. When both coils were bended, the self-inductance decreased to 562 nH and the coupling coefficient changed to 4.2% from 5.0% and the PTE dropped to 37%. The flexible coupled-coils can maintain high PTE even when bended. $L_{s2}$ is 562 nH, and the resonant capacitor $C_{s2}$ thus equals to 27 pF, which is fabricated on-chip. The on-chip capacitance $C_{s2}$ cannot be fabricated with a very exact value and at the same time, is affected by the parasitic capacitor of the loading circuits. One effect of the mismatched $C_{s2}$ is that the LSK scheme may malfunction at the weak coupling condition and robust data transmission will be compromised. In Section III, the issue of LSK scheme is analyzed and a solution is proposed to achieve robust communication.

### III. ANALYSIS OF LSK MODULATION SCHEME

#### A. Ideal LSK Modulation Scheme

We first consider the LSK modulation under ideal situation. Fig. 4 shows an AC voltage source driving the primary LC tank $L_{s1}$ and $C_{s1}$. The secondary LC tank receives the coupled power and delivers to the implant unit. The equivalent loading resistance $R_{Load}$ is 12.5 kΩ, and parasitic capacitor $C_{p2}$ is assumed to be small and can be ignored. During backscattering, the implant unit shunts $R_{SW}$ to modify the loading resistance across the secondary coil. The primary coil input current is given by (4). The primary impedance $Z_{pri}$ and the primary resonator impedance $Z_{11}$ were defined on (5) respectively.

$$I_1(j\omega) = \frac{V_s}{Z_{pri}} \quad (4)$$

$$Z_{pri} = Z_{11} + Z_{eq}, Z_{11} = R_{s1} + j\omega L_{s1} + \frac{1}{j\omega C_{s1}} \quad (5)$$

The equivalent impedance model is shown in Fig. 4(b), and the equivalent impedance $Z_{eq}(j\omega)$ is given by

$$Z_{eq}(j\omega) = \frac{\omega^2 M^2}{R_{s2} + j\omega L_{s2} + 1/j\omega C_{s2} || R_L} \quad (6)$$

$Z_{eq}$ is real at the frequency $\omega_o$.

For low-power application, the load resistor $R_L$ would be large, for example, $R_L > 100\ \Omega$ or even larger than 1 kΩ, so that the factor $L_{s2}/C_{s2}R_L^2$ is much smaller than 1. As discussed in [11], there is only minor advantage in making the equivalent impedance real, while the analysis is made much simpler with good accuracy by designing $L_{s2}$ and $C_{s2}$ to satisfy $\omega_o = \frac{1}{\sqrt{L_{s2}C_{s2}}}$.

We further assume that the output voltage is adequately filtered, which requires $\omega_o C_{s2} R_L \gg 1$. Moreover, $R_L$ is much larger than the parasitic resistor $R_{s2}$. Hence, the imaginary

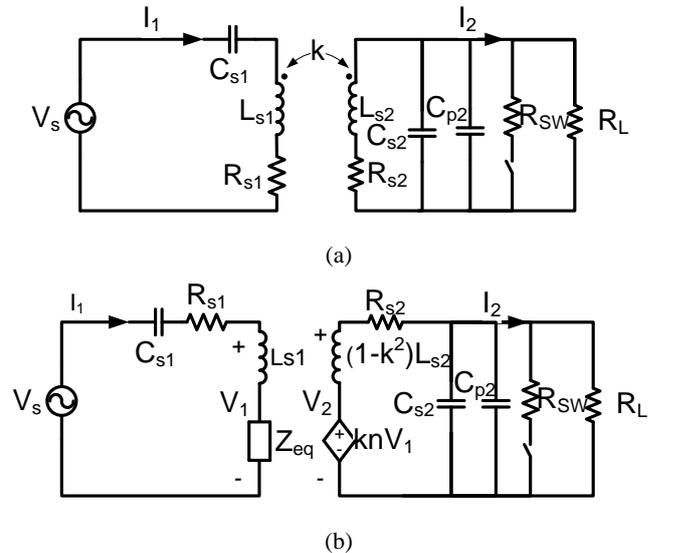

(a)

(b)

Fig. 4. (a) Simplified block diagram of IMDs. (b) Reflected impedance model

part is negligible compared to the real part $R_{eq}$, and $R_{eq}$ is given by

$$R_{eq} \approx \frac{\omega_o^2 k^2 L_{s1} L_{s2}}{(1 + Q_2 Q_L) R_{s2}}, Q_L = \frac{\omega_o L_{s2}}{R_L} \qquad (9)$$

According to (9) $R_{eq}$ increases monotonously w.r.t. $R_L$. Ideally, backscattering changes $R_{eq}$ from $R_L$ to $R_L || R_{SW}$. The equivalent impedance $Z_{eq}$ and $Z_{pri}$ will decrease so $I_1$ will increase. By detecting the amplitude of $I_1$, the transmitter side can decode the uplink signal.

### B. The Secondary LC Tank Mismatch Effect

In practice, it is difficult to make $L_{s2}$, $C_{s2}$ and $R_L$ to satisfy (1). The receiver circuits such as the over-voltage protection (OVP) block and the rectifier will introduce parasitic capacitance $C_{p2}$ to the secondary LC tank. The OVP circuit needs large diodes to sink excessive power, and the rectifier diodes have to be large enough to reduce conduction loss, and both together lead to a large $C_{p2}$. The secondary coil is designed to have high quality factor $Q_2$ to improve PTE, and $L_{s2}$ is designed to be 564 nH and $C_{s2}$ to be 27 pF. From actual measurement, the parasitic capacitance $C_{p2}$ can be up to 12 pF [1], which is not negligible and affects the equivalent impedance.

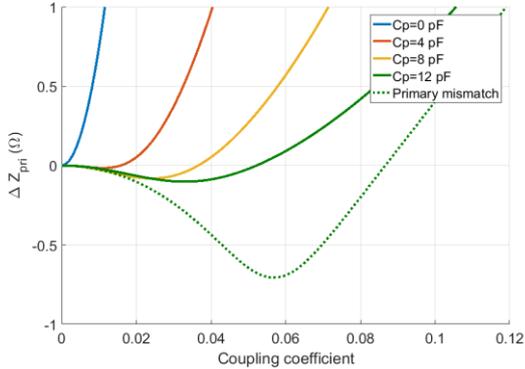

Fig. 5. Primary impedance magnitude difference versus coupling coefficient

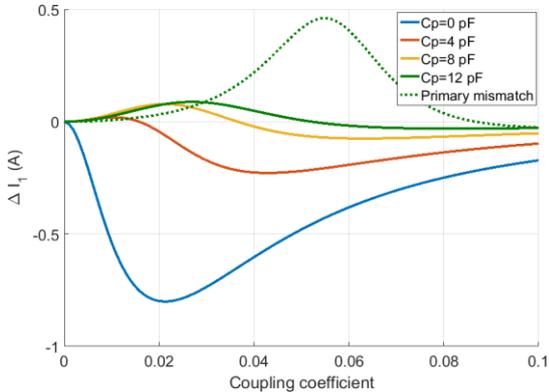

Fig. 6. Primary resonator input current magnitude difference versus coupling coefficient

The equivalent impedance $Z_{eq}(s)$ is shown in (10). $L_{s2}$, $C_{s2}$ and $R_L$ cannot satisfy (1) due to $C_{p2}$. Hence, $Z_{eq}$ is not monotonously changing with $R_L$. Moreover, the primary impedance $Z_{pri}$ is not monotonously increasing with $R_L$. During backscattering, the secondary coil switches from heavy load to light load and the change in the primary impedance $\Delta Z_{pri}$ is given by (7):

Without the parasitic capacitor, $\Delta Z_{pri}$ is a monotonous function of $R_L$, and $\Delta Z_{pri}$ is positive under all coupling

$$\Delta Z_{pri} = Z_{pri_{light}} - Z_{pri_{heavy}} \qquad (7)$$

$$\Delta I_1 = I_{1_{light}} - I_{1_{heavy}} \qquad (8)$$

condition. However, due to $C_{p2}$ of the secondary LC tank, $Z_{pri_H}$ will be smaller than $Z_{pri_L}$ at the weak coupling condition. In the numerical example, the parameter set up were following the structure shown in Fig. 4(a), $L_{s1}$, $L_{s2}$, $R_{s1}$ and $R_{s2}$ were set according to Table I. $C_{s1}$ and $C_{s2}$ were set to satisfy (1). $R_L$ was 12.5 kΩ and $R_{SW}$ was set to 500 Ω. As shown in Fig. 5, when $k$ is lower than 0.05, $\Delta Z_{pri}$ is negative, which means that during backscattering when the secondary loading is switched from light load to heavy load, the amplitude of the primary coil current $I_1$ will decrease instead of increasing, and according to (8) $\Delta I_1$ is positive at weak coupling. The theoretical results were shown in Fig. 6. With the coupling coefficient increasing, $\Delta I_1$ will change from positive to negative, so the uplink signal generated by the LSK scheme will flip consequently.

The situation will be worse if there is also a mismatch in the primary LC tank. We change the simulation set up. The primary resonant capacitance $C_{s1}$ is larger than the designed value by 1% and $C_p$ is still 12 pF. The results of the stimulation are shown in Fig. 5 and Fig. 6, which are labeled as "Primary mismatch". From the stimulation results, it can be seen that due to the primary mismatch, the uplink signal will flip at $k = 0.09$ instead of 0.05, and hence the effective communication distance will further decrease.

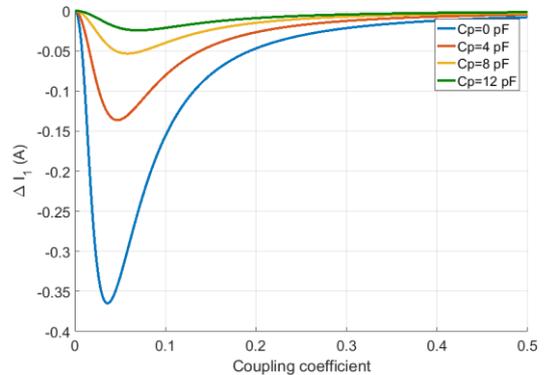

Fig. 7. Detuned primary LC tank input current magnitude difference versus coupling coefficient

$$Z_{eq} = \omega^2 k^2 L_{s1} L_{s2} * \frac{R_L + R_{s2} + \omega^2 * C_2^2 R_L^2 R_{s2} - j\omega(L_{s2} - C_2 R_L^2 + \omega^2 C_2^2 L_{s2} R_L^2)}{(R_L + R_{s2})^2 + \omega^2 [L_{s2}^2 + R_L^2 C_2 (R_{s2}^2 C_2 - 2 L_{s2})]}, C_2 = C_{s2} + C_p \qquad (10)$$

## IV. SIMULATION RESULTS

The theoretical results are shown in Fig. 7. Two cases were simulated by using SPICE to verify the theoretical results. For both cases, $C_p$ was set to 12 pF, and $k$ at 0.06. For the proposed solution, $C_{s1}$ is set at 17.03 pF, which is smaller than the designed value by 1%. The switch was driven by the "SW" signal, and it is active high. When the switch turns on, the secondary loading will change from light load to heavy load, the primary current $I_1$ should increase but as shown in Fig. 8(a), $I_1$ was larger when the secondary was at light load. The flipping issue has been eliminated in Fig. 8(b).

The secondary parasitic capacitance $C_p$ will introduce an negative imaginary part into the equivalent impedance $Z_{eq}$. However, if $C_{s1}$ is made larger than the designed value, $Z_{11}$ will contain a positive imaginary part that could be used to cancel the imaginary part of $Z_{eq}$, so that $\Delta Z_{pri}$ may become negative at weak coupling.

To avoid cancelling between $Z_{eq}$ and $Z_{11}$, the primary resonator could be detuned. By making $C_{s1}$ lower than the designed value, $Z_{11}$ was guaranteed to have a negative imaginary part, such that the uplink signal flipping issue could be eliminated.

## V. CONCLUSIONS

In this work, a polyimide-based flexible coupled-coils was developed for the TsES system, and the inductive link has been modeled in HFSS and fabricated by FPC. The PTE was 39% when the coils were separated by 10 mm. LSK scheme with parasitic parameter has been analyzed and a solution was proposed.

## ACKNOWLEDGMENT

This research is in part sponsored by ….

**Table II Inductive Links Comparison**

|  | [1] | [12] | [13] | This work |
|---|---|---|---|---|
| Link structure | 2-coil | 2-coil | 2-coil | 2-coil |
| Substrate material | FR4 | FR4 | PDMS | Polyimide |
| Operation frequency($Hz$) | 40.68 M | 5 M | 4 M | 40.68 M |
| Relative distance($mm$) | 10 | 10 | 12 | 10 |
| Outer diameter of primary coil($mm$) | 20 | 70 | 37 | 24 |
| Outer diameter of Secondary coil($mm$) | 4.4 | 20 | 16×10 | 6×14 |
| Distance/secondary coil diameter | 2.3 | 0.5 | 1 | 1.6 |
| Sim. or Meas. | Meas. | Meas. | Meas. | Sim. |
| PTE(@ perfect alignment) | 8% | 85.8% | 21% | 39% |

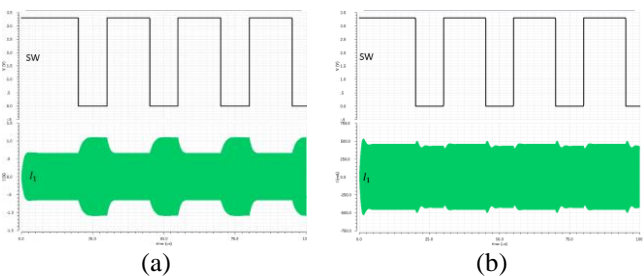

Fig. 8. Simulation results (a) flipping case. (b) potential solution